\documentclass[conference]{IEEEtran}
\IEEEoverridecommandlockouts
\usepackage{url}
\usepackage{cite}
\usepackage{amsmath,amssymb,amsfonts}
\usepackage{algorithmic}
\usepackage{graphicx}
\usepackage{textcomp}
\usepackage{xcolor}
\usepackage[colorinlistoftodos]{todonotes}
\def\BibTeX{{\rm B\kern-.05em{\sc i\kern-.025em b}\kern-.08em
    T\kern-.1667em\lower.7ex\hbox{E}\kern-.125emX}}

\usepackage{color, soul}

\usepackage{tikz}
\usepackage{hyperref}
\usepackage{lipsum}

\newcommand\copyrighttext{%
  \footnotesize \textcopyright 2023 IEEE. Personal use of this material is permitted.
  Permission from IEEE must be obtained for all other uses, in any current or future 
  media, including reprinting/republishing this material for advertising or promotional 
  purposes, creating new collective works, for resale or redistribution to servers or 
  lists, or reuse of any copyrighted component of this work in other works. 
  DOI: \href{https://doi.org/10.1109/ISC257844.2023.10293689}{10.1109/ISC257844.2023.10293689}}
\newcommand\copyrightnotice{%
\begin{tikzpicture}[remember picture,overlay]
\node[anchor=south,yshift=10pt] at (current page.south) {\fbox{\parbox{\dimexpr\textwidth-\fboxsep-\fboxrule\relax}{\copyrighttext}}};
\end{tikzpicture}%
}

\begin{document}

%

\title{Dimensionality Reduction on IoT Monitoring Data of Smart Building for Energy Consumption Forecasting
\thanks{The present work has been developed as part of the SUSTAINABLE project, funded by the European Union’s Horizon 2020 research and innovation program under the Marie Sklodowska-Curie-RISE Grant Agreement No 101007702 https://www.projectsustainable.eu.}
}

\author{\IEEEauthorblockN{Konstantinos Koutras}
\IEEEauthorblockA{\textit{Industrial Systems Institute} \\
\textit{Athena Research Center}\\
Patras, Greece \\
ORCID: 0000-0003-0600-091X}

\\
\IEEEauthorblockN{Athanasios Kalogeras}
\IEEEauthorblockA{\textit{Industrial Systems Institute} \\
\textit{Athena Research Center}\\
Patras, Greece \\
ORCID: 0000-0001-5914-7523}

\and
\IEEEauthorblockN{Agorakis Bompotas}
\IEEEauthorblockA{\textit{Industrial Systems Institute} \\
\textit{Athena Research Center}\\
Patras, Greece \\
ORCID: 0000-0002-6063-8562}

\\
\IEEEauthorblockN{Christos Alexakos}
\IEEEauthorblockA{\textit{Industrial Systems Institute} \\
\textit{Athena Research Center}\\
Patras, Greece \\
ORCID: 0000-0002-8932-6781}

\and 
\IEEEauthorblockN{Constantinos Halkiopoulos}
\IEEEauthorblockA{\textit{Dept. of Management Science and Technology} \\
\textit{University of Patras}\\
Patras, Greece \\
ORCID: 0000-0001-7924-5075}

}
\IEEEpubid{\makebox[\columnwidth]{979-8-3503-9775-8/23/\$31.000~\copyright2023 IEEE \hfill} \hspace{\columnsep}\makebox[\columnwidth]{ }}

\maketitle

\copyrightnotice

\begin{abstract}
The Internet of Things (IoT) plays a major role today in smart building infrastructures, from simple smart-home applications, to more sophisticated industrial type installations. The vast amounts of data generated from relevant systems can be processed in different ways revealing important information. This is especially true in the era of edge computing, when advanced data analysis and decision-making is gradually moving to the edge of the network where devices are generally characterised by low computing resources. In this context, one of the emerging main challenges is related to maintaining data analysis accuracy even with less data that can be efficiently handled by low resource devices. The present work focuses on correlation analysis of data retrieved from a pilot IoT network installation monitoring a small smart office by means of  environmental and energy consumption sensors. The research motivation was to find statistical correlation between the monitoring variables that will allow the use of machine learning (ML) prediction algorithms for energy consumption reducing input parameters. For this to happen, a series of hypothesis tests for the correlation of three different environmental variables with the energy consumption were carried out. A total of ninety tests were performed, thirty for each pair of variables. In these tests, p-values showed the existence of strong or semi-strong correlation with two environmental variables, and of a weak correlation with a third one. Using the proposed methodology, we manage without examining the entire data set to exclude weak correlated variables while keeping the same score of accuracy. 
\end{abstract}


\section{Introduction}

Energy consumption prediction in smart buildings offers numerous benefits that contribute to improved energy efficiency and cost savings \cite{kalogeras2020predictive}. By leveraging advanced data analytics and artificial intelligence, such prediction can anticipate the energy demand of a building over time, optimizing heating, cooling, lighting, and other systems accordingly \cite{cheng2022impact}. This proactive approach enables building managers to fine-tune energy usage, thereby reducing wastage, and lowering overall energy bills. Moreover, energy consumption prediction facilitates better resource planning, allowing occupants to adjust their activities and behaviors to align with expected energy patterns. Over time, this can lead to substantial reductions in greenhouse gas emissions, promoting environmental sustainability. Furthermore, the ability to identify potential energy anomalies or inefficiencies, can lead to smart building corrective actions promptly, ensuring optimal energy performance, and extending the lifespan of equipment \cite{amasyali2018review}.

Energy consumption prediction mostly used technologies comprise data collection from IoT networks, usually related to smart meter and environmental sensors, and forecasting algorithms based on machine \cite{alawadi2020comparison} or deep learning \cite{somu2021deep}. Nevertheless, the energy consumption is not a straightforward task, is highly depended on the sensing capabilities of the devices, the monitored environment (topology, internal architecture, etc.), and the area coverage of the monitoring devices \cite{shah2019review}. Another dimension is the use of Edge and Fog Computing, where low computing resources conduct data analysis to enable rapid reaction to prediction results. 

In such a dynamic environment, the dynamic dimensional reduction of the input is essential to the end of providing a reduction of the complexity of the forecasting algorithms \cite{al2020smart}. The selection of the appropriate input variables / features for the prediction is widely used on many previous works, usually by categorising the variables based on what they are monitoring (i.e. building characteristics, equipment and systems, weather, occupants, and sociological influences) \cite{zhang2021review}. A feature selection, based  on forecasting time series in smart building utilizing the WEKA attributeSelection class, is proposed by González-Vidal et al. \cite{gonzalez2019methodology}. Furthermore, Williams and Short (2020) \cite{williams2020electricity} used Piecewise Aggregate Approximation (PAA) technique for dimensionality reduction, while proposing a forecasting method for electricity demand for decentralised energy management.

The main characteristic of the aforementioned methods is that, in order to proceed to the feature selection and dimension reduction, they require analysis of a large data set. This article presents a novel energy consumption forecasting methodology which is based on the statistical correlation analysis of a relatively small fragment of the historical data set in order to select the appropriate input variables that will provide an accurate prediction.

The rest of the paper is structured as follows. Section 2 presents the methodology followed for the implementation of the dimension reduction technique targeting machine learning algorithms used on a data set retrieved from the IoT network of a real smart building infrastructure, in order to make temporal prediction of the energy consumption. Section 3 presents the experimental evaluation of the proposed methodology, while Section 4 gives a detailed discussion and conclusion about the findings, and the limitations of the approach.

\section{Proposed Methodology}

Developing a model for predicting different continuous values consists mainly of five different parts: the data collection, the pre-processing of the data, the training of a predictive model, the evaluation of the model, and finally, the actual usage of the model in order to make predictions, and aid various decision making processes. This work targets the application of different machine learning and statistical algorithms in order to make predictions of the energy consumption of a smart building. For the prediction of the energy consumption, different measurements from environmental and energy consumption sensors are used. For evaluation purposes, the data is collected from sensors that are a part of a wider IoT network installation in a smart building Living Lab at the premises of the Industrial Systems Institute in Patras, Greece. The environmental parameters which are used are the temperature of the office, the humidity, and lighting levels. Different information and communication technological devices of that office are also connected to smart-plugs, which measure the relevant power associated with their energy consumption. 

For our case-study, data was retrieved from the IoT network platform for a period of one year (June 2020 until end of May 2021). The final data set used for analysis consisted of the variables of the environmental temperature, the humidity, the lighting level, and the energy consumption, with every value been approximately rounded to the minute.

\subsection{Correlation Analysis} 

The pre-processing part of the data set consisted of the usage of inferential statistics, and more specifically, the usage of hypothesis testing of the correlations of the three environmental (temperature, humidity, and lighting level) variables with the energy consumption. From the initial data set, thirty samples of fifty instances were randomly chosen. For every sample of data, and for every Pearson's correlation coefficient (r) between the three pairs of variables of this data (temperature and power, humidity and power, and lighting level and power), there was a correlation t-test performed, with the null hypothesis ($H_0$) being: \textbf{$H_0$:} r = 0, and the alternative hypothesis ($H_1$) being: \textbf{$H_1$:} r $\neq$ 0, on significance level of a $=$ 0.05. A p-value is defined as the probability of obtaining a value more extreme, or equal, than the value that was actually observed. Hence, a t-score with a p-value of $<=0.05$, would be identified as statistically important, making us reject the null hypothesis. As the number of the pairs of variables analysed were three, three t-scores and p-values were acquired from each sample, and ninety overall. Based on this analysis the less significant correlated variables are dismissed from the input.

\subsection{Energy Consumption Forecasting}

Different algorithms were applied in order to predict the energy consumption after five temporal moments, depending on the previous ten values of both the energy consumption and the environmental variables of the temperature, humidity, and lighting level. Since our goals were to determine how our statistical pre-processing affects the prediction outcome and which algorithms can provide accurate predictions on the given problem, we opted to conduct two rounds of experiments using a wide variety of regressors. 

During the first round of experiments, the full data set was used while on the second round we dropped the humidity values because during the pre-processing analysis they displayed little to no correlation to the power consumption values. In both rounds of experiments, the data sets were transformed to the appropriate format to be used by the algorithms, and were split in train-test parts the exact same way to eliminate any differences in performance that could arise from the different handling of the input data. Furthermore, multiple runs of the experiments were carried out in order to eliminate any randomness on the results and the same regression methods were selected at all times. These methods were: Ordinary Least Squares, Ridge Linear Regression, Lasso Linear Regression, K Nearest Neighbors, Decision Tree, Gradient Boosting, Support Vector Machine (SVM) and two different types of Artificial Neural Networks (ANN), one with lots of perceptrons and a simpler one.

Finally, the trained models were employed to predict the power consumption of the test split of the data sets, and their results were evaluated using the most popular regression metrics: Mean Squared Error, Rooted Mean Squared Error, Mean Absolute Error and $R^2$.

\section{Evaluation Use Case }

\subsection{Living Lab Infrastructure} 
\label{ISI LL}

The data retrieved for the purpose of this study was collected from a pilot IoT network installed at the Industrial Systems Institute (ISI) offices in the building of Patras Science Park (PSP), in Patras, Greece. The IoT network sensors where installed in five different areas and were used for environmental parameter and energy consumption monitoring. The technological equipment that was used for energy consumption monitoring consists of a number of different smart-plugs, while the environmental monitoring was accomplished by custom multi-sensor prototype devices based on Arduino platform. Among others, Meazon smart-plugs (\url{https://meazon.com/}) were used for metering the electricity power consumed (in Watt) by devices connected to it. This one variable was the one used in our case for prediction. The custom multi-sensor was mainly used to collect the environmental values of temperature, humidity, and lighting level of an area. For the purpose of this study, one space out of these five mentioned was used to retrieve the data collected by its sensors. This space sensors comprised one Meazon sensor, and one custom Arduino multi-sensor.

The overall IoT system follows the three-tier model suggested by Industrial Internet Reference Architecture (IIRA). At the edge tier, a number of different environmental and energy monitoring sensors resides, which communicate through WiFi, Ethernet or ZigBee, to a central server, using the Message Queuing Telemetry Transport (MQTT) protocol. The platform and the enterprise tier consist of the modules of ThingsBoard, an open-source platform used for IoT networks administration. A Cassandra NoSQL database is also used for storing the telemetries and the device configuration data. More information about the system architecture and the technological procedures for the installation of this IoT network can be found in \cite{alexakos2019building}.

\subsection{Evaluation Data Set}
The ThingsBoard platform that was used for the storage of data collected from the sensors consists of an Application Programming Interface (API) which can provide every value of the stored telemetries you make a query for. The custom Arduino multi-sensor was communicating data for storage, on average, every fifteen seconds, while the Meazon smart-plug did not have any specific time frame for this. Every value that was stored in ThingsBoard, was retrieved for a period of one year, from the 1st of June of 2020 until 31th of May 2021, for both the energy consumption and the environmental monitoring sensor. The data set that was created contained four variables, the environmental temperature of the office been examined, its humidity, the lighting levels, as well as the energy consumption in Watt, metered by the Meazon smart-plug from the different technological equipment and devices connected to it.
\\
As the majority of the instances of the data set consisted of values that were not measured by the two sensors at the same timestamp, further procedures for the appropriate formation of the data were carried out, in order to achieve the creation of a data set with equal timings within the environmental variables and the energy consumption. For this to happen, the data of the environmental variables and their time series were studied separately from those of the energy consumption, and vice versa. For both cases, the timestamps were approximately rounded on a minute, creating a new "artificial timestamp" that the values of the telemetries can be identified. After this step, the energy consumption and the environmental parameter data were merged, creating a new data set with values of all the variables on common timings. In case multiple values were assigned to a specific rounded timestamp, the methodology that was followed was to retain the mean of these values as the one that should be assigned to that specific timestamp. Finally, every instance with a missing value was eliminated, leading to a data set with a total of 372170 records. This data set was the one used for the statistical pre-processing and the implementation of the predictive algorithms.

\subsection{Dimensionality Reduction}

During the first phase of the methodology, we conducted the correlation analysis on the provided data set. An overview of these results can be found in Table \ref{table:pValues}, where the minimum (Min.), the maximum (Max.), and mean of the p-values from the samples analysed, the median, as well as the percentage of the total null hypothesis rejected, are presented.

\begin{table}[h]
\centering
\caption{Results of the Correlation Analysis}
    \begin{tabular}{c||c|c|c|c|c}
         \textbf{Variables} & \textbf{Min.} & \textbf{Max.} & \textbf{Mean} & \textbf{Median} &\textbf{Reject} $\textbf{H}_\textbf{0}$ \\ \hline
         Temperature-Power & 0.002 & 0.749 & 0.092 & 0.035 & 73.3 \%\\
         Humidity-Power & 0.008 & 0.992 & 0.564 & 0.585 & 6.67 \%\\
         Light-Power & 0.001 & 0.941 & 0.312 & 0.195 & 37.67 \%\\         
    \end{tabular}     
\label{table:pValues}
\end{table}

The correlations between the environmental temperature and the energy consumption showed the existence of statistically significant results with p-value $<=$ 0.05 in the majority of the tests performed (22 out of the 30 samples). The correlations between the lighting and the power were statistically significant in almost half of the cases (11 out of the 30 samples), while on the other hand, the correlations between the humidity and the power was not observed to be statistically significant (2 out of the 30 samples). These results indicate that the humidity can be safely removed from the input vector.

\subsection{Evaluation Experiments}

\begin{figure}
\centering
\includegraphics[width=\linewidth]{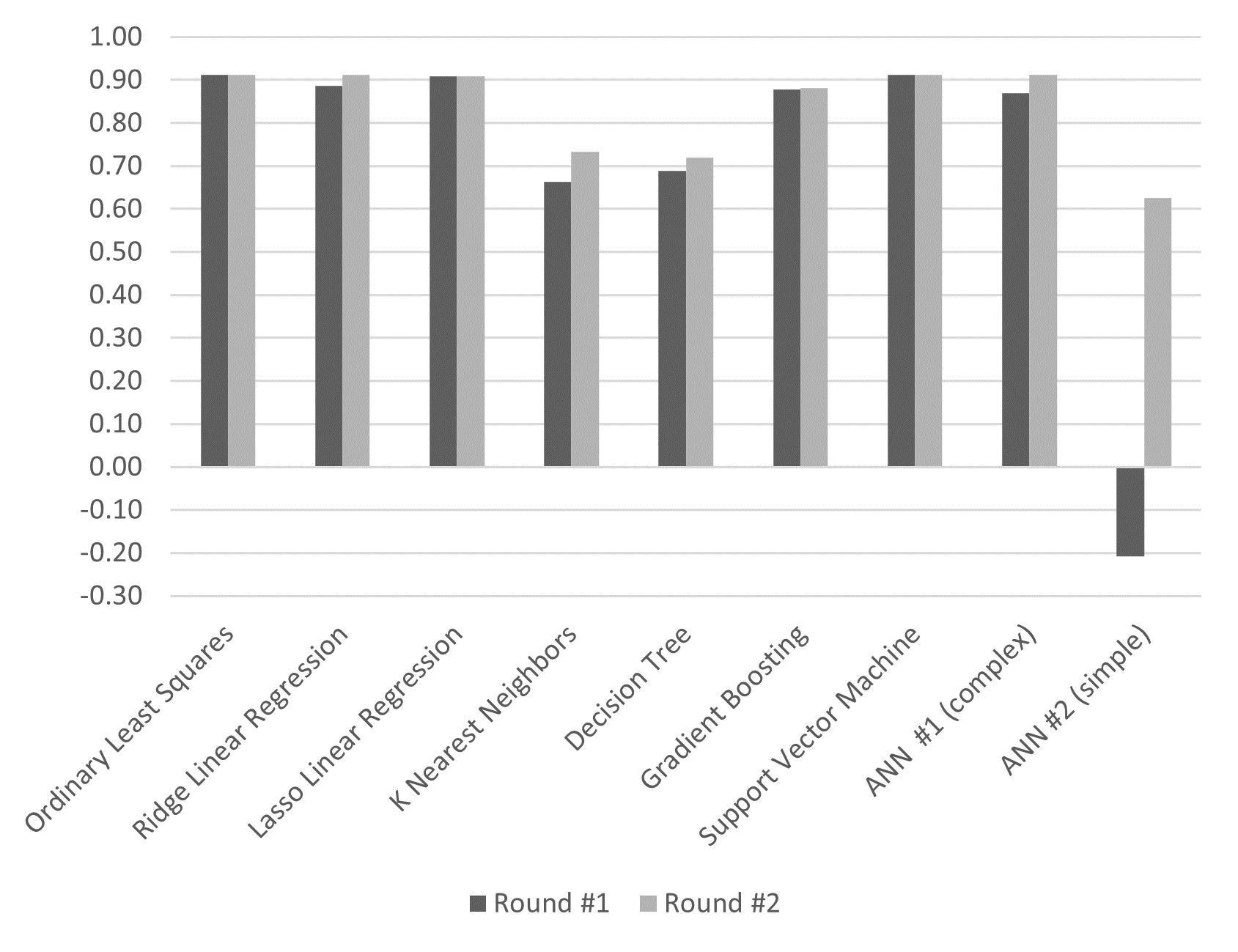}
\caption{$R^2$ comparison of both rounds}
\label{fig:r2Comp}
\end{figure}

\begin{figure}
\centering
\includegraphics[width=\linewidth]{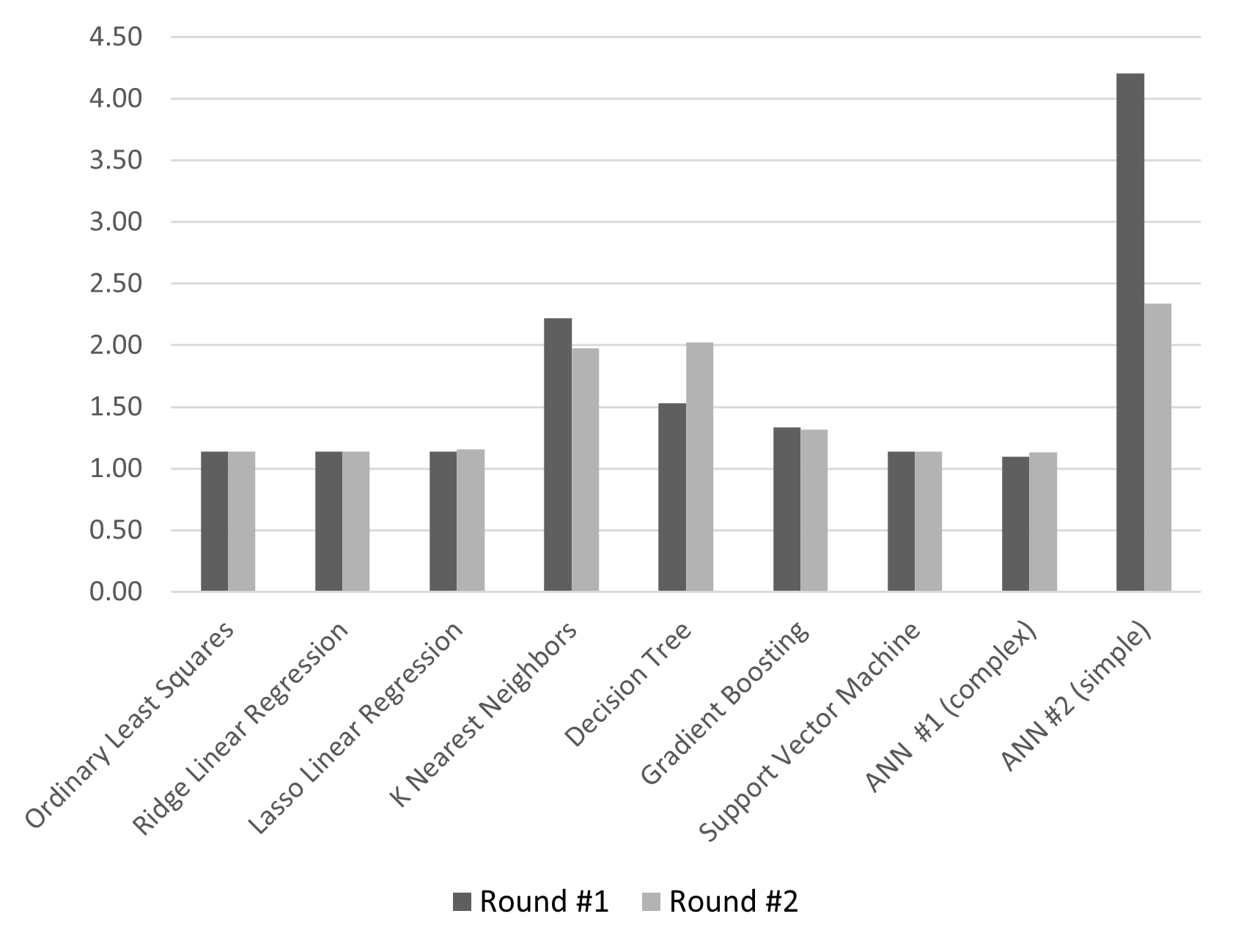}
\caption{RMSE comparison of both rounds}
\label{fig:rmseComp}
\end{figure}

\begin{figure}
\centering
\includegraphics[width=\linewidth]{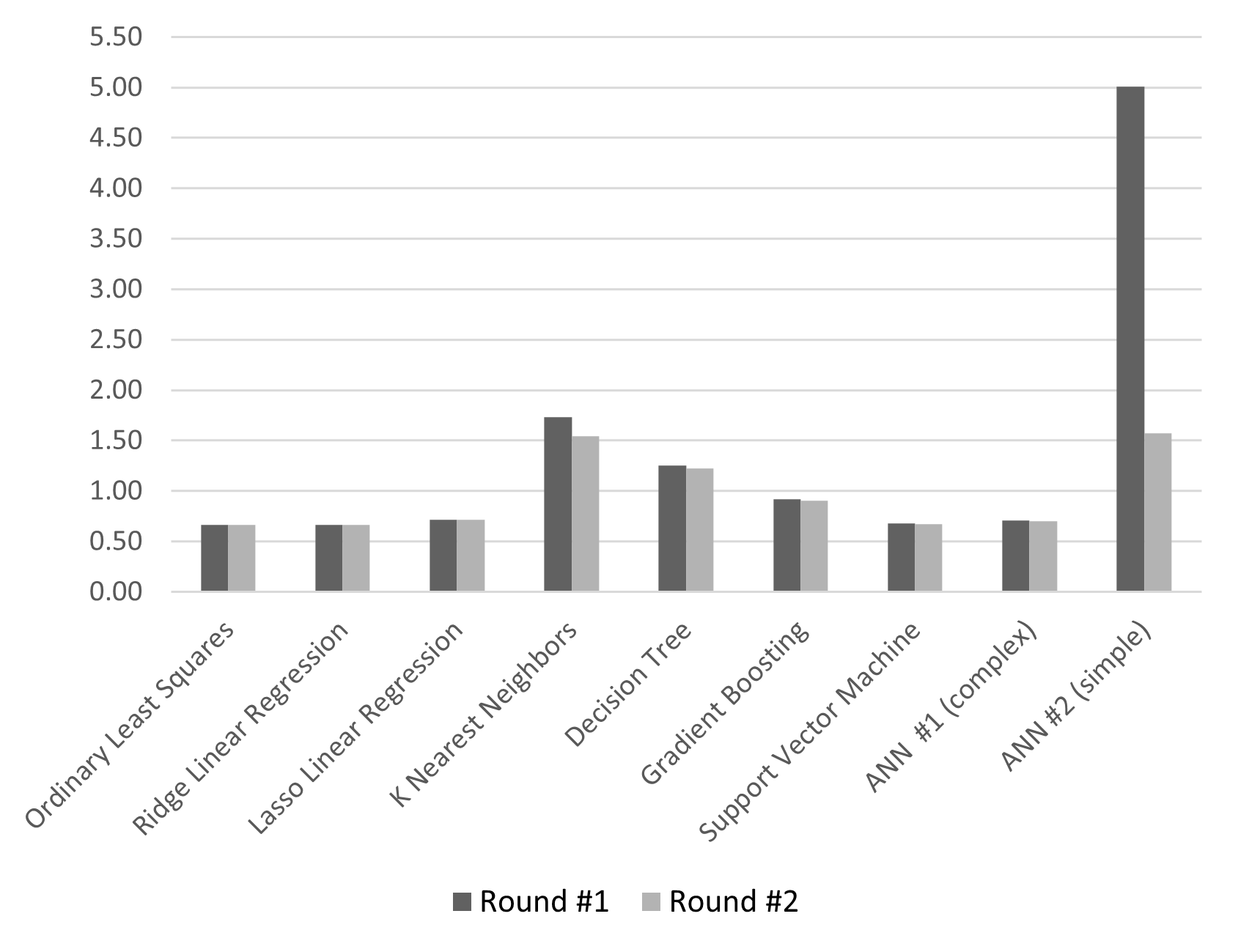}
\caption{MAE comparison of both rounds}
\label{fig:maeComp}
\end{figure}

The next phase included the execution of the two rounds of energy consumption forecasting, one with all the variables and the second with the reduced input set. During both rounds of experiments the algorithms that performed better without the need of excessive fine tuning were the simpler linear ones, namely the Ordinary Least Squares, the Ridge and Lasso regressors. Furthermore, these methods possess the benefits of being much faster to train and have the ability to scale more effortlessly in larger data sets. As can been observed in Tables \ref{table:r1avgs} and \ref{table:r2avgs} and is illustrated in Figures \ref{fig:r2Comp}, \ref{fig:rmseComp} and \ref{fig:maeComp} the K Nearest Neighbors, Decision Tree and the simpler Artificial Neural Network performed poorly in comparison to the other regressors. However, the more complex ANN produced fairly good results as did the Gradient Boosting and the Support Vector Regressor that was one of the best performing algorithms across all the metrics.

\begin{table}[h]
\centering
\caption{Results of the First Round (Avg.)}
    \begin{tabular}{c||c|c|c|c}
        \textbf{Algorithms} & $\textbf{R}^\textbf{2}$ & \textbf{MSE} & \textbf{RMSE} & \textbf{MAE} \\ \hline
        Ordinary Least Squares & 0.9118 & 1.2886 & 1.1352 & 0.6626 \\
        Ridge Linear Regression & 0.8869 & 1.2886 & 1.1352 & 0.6626 \\
        Lasso Linear Regression & 0.9089 & 1.2890 & 1.1384 & 0.7169 \\
        K Nearest Neighbors & 0.6632 & 4.9220 & 2.2185 & 1.7313 \\
        Decision Tree & 0.6891 & 3.3940 & 1.5314 & 1.2516 \\
        Gradient Boosting & 0.8781 & 1.7813 & 1.3346 & 0.9183 \\
        Support Vector Machine & 0.9120 & 1.2958 & 1.1383 & 0.6757 \\
        ANN \#1 (complex) & 0.8692 & 1.2000 & 1.0942 & 0.7083 \\
        ANN \#2 (simple) & -0.2087 & 17.6626 & 4.2027 & 5.0074 \\
    \end{tabular}     
\label{table:r1avgs}
\end{table}

\begin{table}[h]
\centering
\caption{Results of the Second Round (Avg.)}
    \begin{tabular}{c||c|c|c|c}
        \textbf{Algorithms} & $\textbf{R}^\textbf{2}$ & \textbf{MSE} & \textbf{RMSE} & \textbf{MAE} \\ \hline
        Ordinary Least Squares  & 0.9118 & 1.2889 & 1.1353 & 0.6635 \\
        Ridge Linear Regression & 0.9118 & 1.2889 & 1.1353 & 0.6635 \\
        Lasso Linear Regression & 0.9089 & 1.3313 & 1.1538 & 0.7169 \\
        K Nearest Neighbors  & 0.7331 & 3.9000 & 1.9748 & 1.5403 \\
        Decision Tree  & 0.7195 & 4.0992 & 2.0246 & 1.2261 \\
        Gradient Boosting & 0.8814 & 1.7326 & 1.3163 & 0.9045 \\
        Support Vector Machine & 0.9112 & 1.2981 & 1.1393 & 0.6730 \\
        ANN \#1 (complex) & 0.9125 & 1.2782 & 1.1306 & 0.7024 \\
        ANN \#2 (simple) & 0.6259 & 5.4669 & 2.3381 & 1.5697 \\
    \end{tabular}     
\label{table:r2avgs}
\end{table}

Another interesting finding of the conducted series of experiments was the consistency of the aforementioned results. During the multiple repetitions that were executed for both data sets the same algorithms that scored the highest in average were the ones that did so in every iteration without any strange outlier models. As shown in Tables \ref{table:r1devs} and \ref{table:r2devs} the Ordinary Least Squares, the Ridge, the Lasso and the Support Vector regressors performed exceptionally well at this task as the standard deviation between the executions tended to zero. Another very consistent method albeit with not so great predictions was the K Nearest Neighbors that for each run produced almost identical results.

\begin{table}[h]
\centering
\caption{Standard deviations for the First Round}
    \begin{tabular}{c||c|c|c|c}
        \textbf{Algorithms} & $\textbf{R}^\textbf{2}$ & \textbf{MSE} & \textbf{RMSE} & \textbf{MAE} \\ \hline
        Ordinary Least Squares & 0.0000 & 0.0000 & 0.0000 & 0.0000 \\
        Ridge Linear Regression & 0.0664 & 0.0000 & 0.0000 & 0.0000 \\
        Lasso Linear Regression & 0.0000 & 0.1125 & 0.0534 & 0.0000 \\
        K Nearest Neighbors & 0.0000 & 0.0000 & 0.0000 & 0.0000 \\
        Decision Tree & 0.0112 & 1.9327 & 0.6509 & 0.0172 \\
        Gradient Boosting & 0.0000 & 0.0003 & 0.0001 & 0.0001 \\
        Support Vector Machine & 0.0025 & 0.0000 & 0.0000 & 0.0000 \\
        ANN \#1 (complex) & 0.1019 & 0.0218 & 0.0099 & 0.1446 \\
        ANN \#2 (simple) & 0.0647 & 0.9455 & 0.1119 & 5.1609 \\
    \end{tabular}     
\label{table:r1devs}
\end{table}

\begin{table}[h]
\centering
\caption{Standard deviations for the Second Round}
    \begin{tabular}{c||c|c|c|c}
        \textbf{Algorithms} & $\textbf{R}^\textbf{2}$ & \textbf{MSE} & \textbf{RMSE} & \textbf{MAE} \\ \hline
        Ordinary Least Squares & 0.0000 & 0.0000 & 0.0000 & 0.0000 \\
        Ridge Linear Regression & 0.0000 & 0.0000 & 0.0000 & 0.0000 \\
        Lasso Linear Regression & 0.0000 & 0.0000 & 0.0000 & 0.0000 \\
        K Nearest Neighbors & 0.0000 & 0.0000 & 0.0000 & 0.0000 \\
        Decision Tree & 0.0107 & 0.1558 & 0.0387 & 0.0138 \\
        Gradient Boosting & 0.0000 & 0.0000 & 0.0000 & 0.0000 \\
        Support Vector Machine & 0.0000 & 0.0000 & 0.0000 & 0.0000 \\
        ANN \#1 (complex) & 0.0072 & 0.1048 & 0.0464 & 0.0902 \\
        ANN \#2 (simple) & 0.0078 & 0.113829 & 0.0243 & 0.0501 \\
    \end{tabular}     
\label{table:r2devs}
\end{table}

\begin{figure}
\centering
\includegraphics[width=\linewidth]{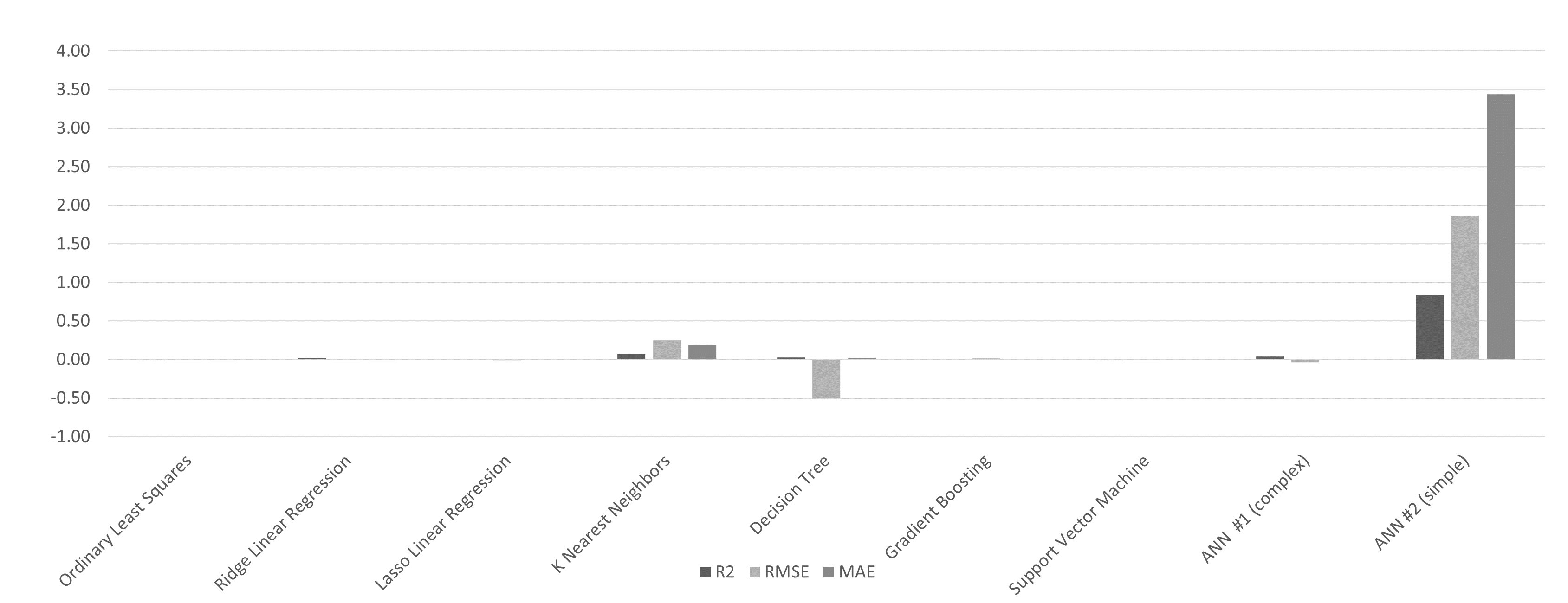}
\caption{Absolute differences between the two rounds (Avg.)}
\label{fig:rndsComp}
\end{figure}

Lastly, one of our most interesting observations during this experimentation process was that the removal of the humidity variable, that was proved to be uncorrelated with the power consumption during the statistical analysis in the pre-processing phase, benefited not only the time needed for fitting our models but also their accuracy. This is evident in Table \ref{table:r1r2avgs} that contains the subtraction of the average scores of the two runs (positive values mean that the reduced data set outperformed the full one). As also depicted graphically in Figure \ref{fig:rndsComp}, the differences between the algorithms that produced the best predictions are trivial but there was a huge improvement in the results of the simpler ANN. Moreover, the removal of the humidity factor also enhanced the consistency of all the trained models as is demonstrated in Table \ref{table:r1r2stds} and Figure \ref{table:r1r2stds}.

\begin{table}[h]
\centering
\caption{Absolute differences between the two rounds (Avg.)}
    \begin{tabular}{c||c|c|c|c}
        \textbf{Algorithms} & $\textbf{R}^\textbf{2}$ & \textbf{MSE} & \textbf{RMSE} & \textbf{MAE} \\ \hline
        Ordinary Least Squares & 0.0000 & -0.0003 & -0.0001 & -0.0009 \\
        Ridge Linear Regression & 0.0249 & -0.0003 & -0.0001 & -0.0009 \\
        Lasso Linear Regression & 0.0000 & -0.0422 & -0.0185 & 0.0000 \\
        K Nearest Neighbors & 0.0699 & 1.0220 & 0.2437 & 0.1911 \\
        Decision Tree & 0.0304 & -0.7052 & -0.1824 & 0.0255 \\
        Gradient Boosting & 0.0033 & 0.0488 & 0.0184 & 0.0138 \\
        Support Vector Machine & -0.0008 & -0.0023 & -0.0010 & 0.0027 \\
        ANN \#1 (complex) & 0.0433 & -0.0782 & -0.0351 & 0.0059 \\
        ANN \#2 (simple) & 0.8346 & 12.1957 & 1.8646 & 3.4377 \\
    \end{tabular}     
\label{table:r1r2avgs}
\end{table}

\begin{figure}
\centering
\includegraphics[width=\linewidth]{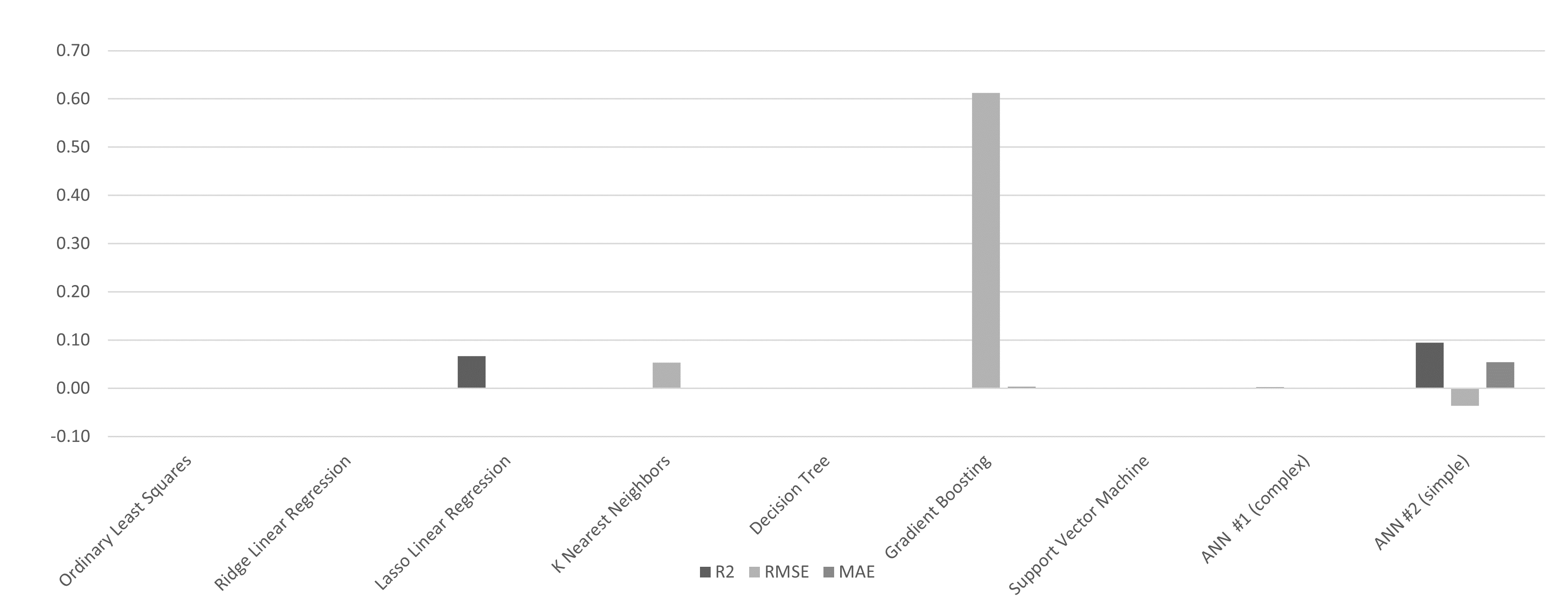}
\caption{Absolute differences between the two rounds (Std. Deviation)}
\label{fig:rndsComp2}
\end{figure}

\begin{table}[h]
\centering
\caption{Absolute differences between the two rounds (Std. Deviation)}
    \begin{tabular}{c||c|c|c|c}
        \textbf{Algorithms} & $\textbf{R}^\textbf{2}$ & \textbf{MSE} & \textbf{RMSE} & \textbf{MAE} \\ \hline
        Ordinary Least Squares & 0.0000 & 0.0000 & 0.0000 & 0.0000 \\
        Ridge Linear Regression & 0.0000 & 0.0000 & 0.0000 & 0.0000 \\
        Lasso Linear Regression & 0.0664 & 0.0000 & 0.0000 & 0.0000 \\
        K Nearest Neighbors & 0.0000 & 0.1126 & 0.0534 & 0.0000 \\
        Decision Tree & 0.0000 & 0.0000 & 0.0000 & 0.0000 \\
        Gradient Boosting & 0.0006 & 1.7770 & 0.6123 & 0.0034 \\
        Support Vector Machine & 0.0000 & 0.0003 & 0.0001 & 0.0001 \\
        ANN \#1 (complex) & 0.0025 & 0.0000 & 0.0000 & 0.0000 \\
        ANN \#2 (simple) & 0.0948 & -0.0830 & -0.0365 & 0.0543 \\
    \end{tabular}     
\label{table:r1r2stds}
\end{table}


\section{Conclusion}

During our experiments we tried to measure the effectiveness of the proposed pre-processing method that employs statistical analysis and hypothesis testing to discover strong correlations that can aid the creation of more concise data sets when dealing with regression problems in time series. During the experimentation process a series of algorithms were tested and were executed multiple times. The models that were trained using the reduced data set that was produced thanks to this pre-processing method were able to either match the accuracy or greatly surpass the accuracy of those that were trained on the full data set. Furthermore, the training on the pre-processed data set yielded in each repetition of the experiment models that had little difference between them. This consistency is a sought out property of every ML pipeline as the experimental results can be easily recreated in a production environment. Last but not least, the time required for fitting a model in the reduced data set was significantly less than what was required for the full one especially for the non-linear regressors such as the SVM and ANNs. \\
Limitations associated with the present work can be mainly classified into two categories. The first has to do with the usage of data from a multitude of sensors increasing complexity and enabling prediction of energy consumption. Secondly, synchronization of all sensors would enable to communicate telemetries in the same time frames, mandating no pre-processing for the rounding of timestamps that we enforced in this work. Further to these aspects, future research could also focus on how different parameters such as the season, or the different time frames during a day, can affect the results of the predictive models. The work presented in this paper could be extended to other domains as well, such as deployed sensory infrastructures in the context of smart and circular cities, or in the context of precision agriculture and agriculture 4.0 \cite{alexopoulos2023complementary}. 

\bibliographystyle{IEEEtran}
\bibliography{WSACC23_Koutras.bib}

\end{document}